\documentclass[11pt,aps,prd,preprint,nofootinbib]{revtex4}
\usepackage{graphicx,amsmath,amssymb}
\usepackage[usenames,dvipsnames]{pstricks}
\usepackage{epsfig}

\begin{document}

\title{Parameter Independence and Outcome Independence \\
in Dynamical Collapse Theories}
\author{Travis Norsen}
\affiliation{Marlboro College \\ Marlboro, VT  05344
\\
\vspace{.5 cm}
}

\date{October 1, 2010
\\
\vspace{1 cm}}

\begin{abstract}
Ghirardi, Grassi, Butterfield, and Fleming have previously argued 
that two distinct formulations of dynamical reduction (GRW) theories 
have distinct non-local properties: one (the ``non-linear'' formulation)
violates Parameter Independence  while the other (the ``linear +
cooking'' formulation) violates Outcome Independence.  The claim is
then that, since theories violating Outcome Independence are
compatible with relativity in a way that theories violating Parameter
Independence are not, the ``linear + cooking'' dynamical reduction
theories hold the key to reconciling quantum non-locality with
relativity.   I review and
assess this claim, arguing that, while ultimately misleading, there is
an interesting and important lesson to be extracted.
\end{abstract}

\maketitle

\section{Introduction}

In an earlier paper \cite{bvj} I criticized Jon Jarrett's distinction
\cite{jarrett} between (what Shimony \cite{shimony} later dubbed)
``Parameter Independence'' (PI) and ``Outcome Independence'' (OI).
Jarrett had shown that Bell's local causality condition \cite{bell}
was equivalent to the conjunction of PI and OI, and he argued that PI
\emph{alone} captured relativity's alleged prohibition on superluminal
causation.  This suggested in turn that candidate quantum theories
violating OI (such as orthodox QM and the GRW spontaneous collapse
theory) should be preferred over theories violating PI (such as the de
Broglie - Bohm pilot-wave theory) because of the former's 
favorable compatibility with relativity theory.  It was brought out in
Ref. \cite{bvj}, however, that Jarrett's argument was based on a
simple confusion about Bell's local causality condition; correcting
the confusion clarified that violations of OI and PI are
equally troubling from the point of view of relativity, suggesting in
turn that the very distinction between OI and PI has no useful role to
play in guiding ongoing attempts to reconcile quantum theory and
relativity.  

In subsequent private communication, however, GianCarlo
Ghirardi pointed out to me two papers \cite{GGBF, GGBF2}
he had coauthored with R. Grassi,
J. Butterfield, and G. Fleming in which the authors (GGBF) show that
two different mathematical formulations of the GRW-type dynamical
collapse theory (in the context of non-relativistic quantum mechanics)
have distinct non-local properties:  one (the ``non-linear''
formulation) violates PI while the other (the ``linear + cooking''
formulation) violates OI. 

The alleged implication was that, because -- according to Jarrett and,
by a rather different argument, also Ghirardi and Grassi
\cite{ggappraisal} -- theories
violating OI are supposed to be more compatible with relativity than
theories violating PI, one should preferentially pursue ``linear + 
cooking'' theories in the quest for a genuinely relativistic quantum
theory without observers.  My intuition was that this could not
possibly be correct, for three reasons:  first, Jarrett's argument for
the relevant premise (that theories violating OI are more compatible
with relativity than theories violating PI) is simply mistaken; 
second, the
``non-linear'' and ``linear + cooking'' formulations of GRW seem
somehow obviously to be just two mathematically distinct ways of 
presenting the same theory, making it implausible that they should 
exhibit meaningfully different types of nonlocality; and third, 
since the publication by GGBF, a genuinely relativistic GRW-type 
theory has been produced \cite{rGRWf} and its formulation is 
explicitly ``non-linear'' (as opposed to ``linear + cooking'').
(Though its author, Roderich Tumulka, 
assures me it could also be formulated in ``linear
+ cooking'' terms.)

Sorting all of this out, however, turned up some interesting points.
The purpose of this note is simply to share them.

\section{Two formulations of GRW}

The main motivation for the GRW-type dynamical collapse theories 
is to solve
the measurement problem.  Where orthodox QM contains two distinct
rules for the time-evolution of state vectors (the 
Schr\"odinger equation and the collapse rule) and so involves an
allegedly fundamental (but in fact vague and ill-defined) distinction
between processes that are ``measurements'' and those that aren't, GRW
incorporates a stochastic collapse process into the Schr\"odinger
equation itself, thus allowing a single uniform dynamics for the wave
function which applies all the time (regardless of whether a
``measurement'' is happening or not).  See Ref. \cite{GRW} for 
the original proposal, and Ref. \cite{GRWSEP} for a recent review.
Ref. \cite{irremediably} contains further discussion, partially in
response to the already-cited Ref. \cite{bvj}, of the prospects of
reconciling quantum non-locality with relativity within a dynamical
collapse picture.\footnote{Perhaps I can also take the opportunity here to
  clarify a misunderstanding that runs through
  Ref. \cite{irremediably}, based on some sloppiness in my own earlier
  paper.  In Ref. \cite{bvj}, my aim was principally to refute
  Jarrett's claim that violation of Bell's local causality condition
  didn't necessary entail the existence of any superluminal causation
  -- i.e., to refute Jarrett's claim that Bell's local causality
  condition was somehow too strong as an attempt to capture
  relativity's (alleged) prohibition on superluminal causation.  I
  thus largely took for granted what Jarrett himself had taken for
  granted:  that relativity does indeed genuinely prohibit causal
  influences between spacelike separated events.  In fact, though, the
  connection between superluminal causation and relativity is
  more subtle.  They are not necessarily incompatible, and I agree 
  entirely with Ghirardi that, for example, Tumulka's recent theory
  \cite{rGRWf} provides a very promising concrete example of how they
  can indeed be reconciled within a GRW framework.  To summarize,
  where Ghirardi not unreasonably reads Ref. \cite{bvj} as claiming an
  ``irremediable conflict'' between the empirically correct
  predictions of quantum theory and
  relativity as such, I in fact intended a rather different claim:  
  that there
  is an ``irremediable conflict'' between the empirically correct
  predictions of quantum theory and the idea of exclusively
  sub-luminal causation that many people -- \emph{reasonably, but 
  perhaps in the end wrongly} --
  have taken to be a requirement of relativity.}

We will be specifically concerned with the application of this theory
to the usual kind of EPR-Bell situation.  Suppose in particular that 
a pair of
spin-1/2 particles in the spin singlet state is produced by a source,
with each particle subsequently flying toward spatially-separated
detectors on the left (L) and right (R) where they are subjected (or
perhaps not subjected) to measurements of some spin component.  For
our purposes it will suffice to consider the same restricted situation
considered by GGBF:  the experimenter on the
left always measures his particle along a fixed direction,
while the experimenter on the right chooses either to make a
measurement along that same fixed direction (which we denote 
$\hat{a}$) or to make no measurement at all (which we denote $*$).  

For this situation, Parameter Independence is the requirement that,
for example, the probability assigned to the outcome ``up'' for the
experiment on the left, when conditioned on the initial state of the
particle pair (and any other relevant ``hidden variables''), should be
independent of whether the experimenter on the right chooses $\hat{a}$
or $*$.  That is, PI is the requirement that
\begin{equation}
P_{\lambda}( \text{up}_L | \hat{a} ) = P_{\lambda}
(\text{up}_L | *).
\label{pi}
\end{equation}
By contrast, Outcome Independence (OI) is the requirement that, when
the setting $s_R \in \{\hat{a}, * \}$ of the distant apparatus is 
specified, the probability for
a specific outcome on the left doesn't depend on the outcome $O_R \in
\{\text{up}_R, \text{down}_R\}$
on the right.  For example:
\begin{equation}
P_{\lambda}(\text{up}_L | s_R, O_R) = P_{\lambda}(\text{up}_L | s_R).
\end{equation}
Jarrett showed, correctly, that Bell's local causality condition (that
is, the condition we know now with reasonable certainty, based on
experiments, to be false) is logically equivalent to the conjunction
of PI and OI.  Hence, empirically adequate theories must violate
either PI or OI (at least if one allows freely-operating
experimenters, sometimes called the ``no conspiracy'' assumption).
(Jarrett was wrong, however, to think that theories
violating PI necessarily contain superluminal causation, while those
violating OI don't.  See Ref. \cite{bvj})

It will also be important later to appreciate
that, for deterministic theories, the outcome $O_R$ is necessarily a
function of $\lambda$ and the apparatus settings, so its specification is
necessarily redundant (once these have been given).  Hence,
deterministic theories cannot violate OI and so must (if they violate
Bell's local causality condition, i.e., if they are empirically
viable) violate PI.  

In the following two subsections, I briefly sketch a highly simplified
version of each of the (allegedly distinct) formulations of GRW.  In
the next section I then discuss
how each relates to OI and PI.  See Ref. \cite{GGBF} 
for more details on how the
``toy model'' versions sketched here relate to the real GRW theory.
The basic idea is to consider explicitly only the degrees of freedom
associated with the entangled particle pair, and to simply assume the
occurrence of a collapse when a measurement is made.  This perhaps
looks suspiciously like ordinary QM (with its distinct dynamical
evolution laws for measurement and non-measurement situations), but it
should be clear in the present context how the talk of
collapses-upon-measurements should be understood from the GRW
perspective.  Note also that, to make the relevant conceptual points
as clear as possible, various probabilities have been rounded off:
for example, events which might occur (with tiny but nonzero
probability) according to the real GRW theories are here assigned
probability zero.  One should therefore understand the relevant
equations as involving only approximate equalities.

\subsection{The ``non-linear'' formulation}

GRW is a stochastic theory.  For the purpose of indicating how the 
theory accounts for the experiment sketched above, it will
suffice to mock up the stochastic aspect  with a pair of
coin flips -- one on the right and one on the left.  (This
simplifying picture is adapted from discussions with Daniel Bedingham
based on Ref. \cite{bedingham}.)
There is
then a certain (non-linear) algorithm through which the incoming state
vector, together with the results of the appropriate coin flips,
determines how the state vector evolves through the measurements.
Although the algorithm in principle is rather complicated -- it is
nothing but the (non-linear, stochastic) evolution equation that
replaces the Schr\"odinger equation for GRW -- we can present here
a much simplified algorithm that captures (to the accuracy described
above and for the limited purpose of analyzing the simple EPR
experiment at hand) all the relevant aspects of the theory. 

Here is the algorithm.  For a given measurement, a certain
quantum state $|\psi_{in} \rangle$  will ``enter'' and interact 
with the measuring device.  After the measurement, the device will
register either ``up'' or ``down'' -- and the quantum state will have
collapsed to the corresponding eigenstate of the appropriate spin
operator -- as follows:
\begin{itemize}

\item  If $\| \hat{P}_{\uparrow} |\psi_{in} \rangle \|^2 \gg \|
  \hat{P}_{\downarrow} |\psi_{in}\rangle \|^2$ then the device will
  register the result ``up'' and the quantum state will collapse to
  $|\psi_{out} \rangle = P_{\uparrow} |\psi_{in}\rangle / \|
  P_{\uparrow} |\psi_{in} \rangle \|$.

\item  If $\| \hat{P}_{\uparrow} |\psi_{in} \rangle \|^2 \ll \|
  \hat{P}_{\downarrow} |\psi_{in}\rangle \|^2$ then the device will
  register the result ``down'' and the quantum state will collapse to
  $|\psi_{out} \rangle = P_{\downarrow} |\psi_{in}\rangle / \|
  P_{\downarrow} |\psi_{in} \rangle \|$.

\item If $\| \hat{P}_{\uparrow} |\psi_{in} \rangle \|^2 \approx \|
  \hat{P}_{\downarrow} |\psi_{in} \rangle \|^2$ then the device will
  register the result ``up'' or ``down'' (and the state will collapse
  in the associated way) according to whether the coin flip comes up
  ``heads'' or ``tails''.  

\end{itemize}
Here, the projection operators $\hat{P}_{\uparrow}$ and
$\hat{P}_{\downarrow}$ onto the spin-up and spin-down states
(respectively) should be understood as acting on the degrees
of freedom associated with whichever particle is being measured 
(L or R).  

Notice that the coin flip is only used, so to speak, as a tie-breaker.
If the incoming quantum state is already dominantly spin-up or
spin-down, the evolution simply preserves this property (and the
measuring device reveals it, and the coin flip is irrelevant).  But 
if the incoming quantum state is a balanced superposition of spin-up 
and spin-down, then the coin flip determines the outcome and the
quantum state collapses accordingly.  

Notice also that the algorithm is both stochastic -- the
outcome of the measurement is (in the general case) influenced by the 
result of the coin flip -- and non-linear:  the algorithm itself
``looks at'' the incoming state vector in order to decide how (if
at all) the state vector should evolve through the course of the
measurement interaction.  

Note finally that, in the event that no measurement occurs, there is
by definition no interaction between the particle pair and the
measuring device, in which case $|\psi_{out}\rangle$ will necessarily
just equal $|\psi_{in}\rangle$.  

Let's now see how this works in the EPR-Bell scenario at hand.  We
consider a Lorentz frame in which the experiment on the right happens
first.  
Then the quantum state $| \psi_{in}^R \rangle$ which enters the device 
on the right is just the initial singlet state:
\begin{equation}
| \psi_0 \rangle = \frac{1}{\sqrt{2}} \left( 
| \uparrow L, \downarrow R \rangle - | \downarrow L, \uparrow R \rangle
\right). 
\end{equation}
In the case that the experimenter on the right chooses to actually
make a measurement, the algorithm sketched above will necessarily go
to the tie-breaker -- that is, the outcome (``up'' or ``down'') will
be determined by the result of the coin flip ($H_R$ or $T_R$,
respectively) and the quantum state will collapse accordingly.  Thus,
the quantum state which subsequently enters the 
device on the left (ignoring the overall pure phases which are
irrelevant here) will be given by
\begin{equation}
|\psi_{in}^L \rangle = \left\{ 
\begin{array}{lll}
|\downarrow L, \uparrow R \rangle & \text{if} & H_R \\
|\uparrow L, \downarrow R \rangle & \text{if} & T_R \\
\end{array}
\right.
\label{psiinL}
\end{equation}
It then follows from the above algorithm that the coin flip on the
left will necessarily be irrelevant to the outcome there:  the outcome
on the left will be ``up'' or ``down'' according to whether the state
vector is already (respectively) a spin-up or spin-down eigenstate, 
i.e., according to whether (respectively) $T_L$ or $H_L$ was realized
on the right.

This scheme thus 
clearly reproduces the familiar quantum mechanical statistics for both
the individual and joint outcomes of the two experiments.

Now we must also consider the possibility that the experimenter on the
right chooses instead not to make a measurement at all.  In that
event, no collapse occurs prior to the measurement on the left and so:
\begin{equation}
| \psi_{in}^L \rangle = | \psi_0 \rangle .
\end{equation}
And this of course implies, again according to the above algorithm, 
that the coin flip on the left will now determine the outcome there,
with the outcome being ``up'' if the result is $H_L$ and ``down'' if
the result is $T_L$.  And so 
the scheme reproduces the correct statistics for this scenario, too.

Though both simplified and schematic (even relative to the already
simplified and schematic version in GGBF) this model
accurately captures all the relevant aspects of the ``non-linear'' GRW
theory's explanation for the measurement statistics in this kind of experiment.

\subsection{The ``linear + cooking'' formulation}

Let us now turn to the alternative ``linear + cooking'' formulation of 
GRW.  The idea here is to replace the explicitly non-linear (stochastic)
dynamics of the previous formulation, with a fully linear dynamics
that is then supplemented by some post-hoc readjustments of the
probabilities.  As before, we will explain the dynamics by tracing 
the evolution of the quantum 
state vector through its interaction with the two measuring devices,
assuming a Lorentz frame in which the device on the right is
encountered first.  And as before, the stochastic element of the
theory will be captured by a pair of 
coin flips, one on each side.  The main difference with the ``non-linear''
formulation will be that here the coin flip will \emph{always} determine the
measurement outcome and the state to which the incoming state
collapses (as opposed to being used only in the event of a tie).  

Let's again see how this works by following the evolution through the
two measurements, starting with the case that the experimenter on the
right does decide to make a measurement.  
As indicated, the outcome of the experiment is
determined by an un-biased coin flip, with $H_R$ producing a result
``$\text{up}_R$'' and $T_R$ producing a result ``$\text{down}_R$''.  
In addition, the state
vector collapses, though (for reasons that will become clear shortly)
we will here impose the collapse by acting with the appropriate
projection operator -- but \emph{not} re-normalizing the state.  Thus,
the quantum state later entering the device on the left will be given
by 
\begin{equation}
|\psi_{in}^L \rangle = \left\{ 
\begin{array}{lll}
-|\downarrow L, \uparrow R \rangle / \sqrt{2} & \text{if} & H_R \\
|\uparrow L, \downarrow R \rangle / \sqrt{2} & \text{if} & T_R \\
\end{array}
\right. .
\label{psiinL}
\end{equation}
The outcome on the left is then also determined by an unbiased coin
flip, with $H_L$ indicating ``$\text{up}_L$'' and $T_L$ indicating
``$\text{down}_L$''.   It
is also necessary now to consider the further state vector collapse
associated with the measurement on the left:  the final
quantum state (after interaction with both measuring devices, but
without imposing a re-normalization after the collapses) is
\begin{equation}
|\psi_{final}\rangle = 
\frac{1}{\sqrt{2}} \left\{
\begin{array}{lll}
\hat{P}_{\uparrow L} \hat{P}_{\uparrow R} |\psi_0 \rangle 
 = |\emptyset \rangle & \text{if} & H_L \, H_R \\
\hat{P}_{\uparrow L} \hat{P}_{\downarrow R} |\psi_0 \rangle 
 = |\uparrow L, \downarrow R \rangle & \text{if} & H_L \, T_R \\
\hat{P}_{\downarrow L} \hat{P}_{\uparrow R} |\psi_0 \rangle 
 = -|\downarrow L, \uparrow R \rangle & \text{if} & T_L \, H_R \\
\hat{P}_{\downarrow L} \hat{P}_{\downarrow R} |\psi_0 \rangle 
 = |\emptyset \rangle & \text{if} & T_L \, T_R \\
\end{array}
\right.
\end{equation}
where $|\emptyset \rangle$ is the zero vector.  The two coin flips were
un-biased \emph{and independent}, so each of the four possible joint
outcomes ($\{H_L, H_R\}$, $\{H_L, T_R\}$, $\{T_L,H_R\}$, and
$\{T_L,T_R\}$)  have equal 
probabilities of 1/4,
as (consequently) do each of the four possible joint outcomes of the
experiments (``up on the left and up on the right'', and so on). 

These, however, represent only the ``raw'' probabilities.  We now
readjust these joint probabilities -- i.e., ``cook'' -- by 
re-weighting the probability of each joint outcome according to the
squared modulus of the associated final quantum state.  Two of these
possible final states have zero norm, and the other two have equal
norms, so the upshot of the ``cooking'' is the assertion that, at the
end of the day,
\begin{eqnarray}
P_{cook}(\text{up}_L, \text{up}_R) & = & 0 \nonumber \\
P_{cook}(\text{up}_L, \text{down}_R) & = & 1/2 \nonumber \\
P_{cook}(\text{down}_L, \text{up}_R) & = & 1/2 \nonumber \\
P_{cook}(\text{down}_L, \text{down}_R) & = & 0 \nonumber  
\end{eqnarray}
so that we again reproduce the usual quantum mechanical statistics for
both the individual and joint outcomes.  

Now let us finally walk through the ``linear + cooking'' formulation in the
case that the experimenter on the right chooses not to make any
measurement.  The only difference is that (since there is no coupling
between the experimental apparatus on the right and the particle pair)
the state of the pair which enters the apparatus on the left is the
original singlet state.  There are then only two possible final state
vectors -- the same two which $|\psi_{in}^L \rangle$ might have been
had the experimenter on the right made a measurement -- and they have
equal norms.  So the ``raw'' and ``cooked'' probabilities turn out to
be equal:
\begin{eqnarray}
P_{cook}(\text{up}_L) &=& 1/2 \nonumber \\
P_{cook}(\text{down}_L) &=& 1/2 \nonumber \\
\end{eqnarray}
and we thus reproduce the expected 50/50 chances for the two
possible outcomes on the left.

\section{Equivalence and non-locality}

Not too much analysis is necessary to make it clear that the
``non-linear'' and ``linear + cooking'' formulations are completely
equivalent.  In the event that the experimenter on the right makes a
measurement, the predictions of the ``non-linear'' formulation come
down to the following set of probability statements:
\begin{eqnarray}
P(\text{up}_R) = 1/2 \nonumber \\
P(\text{down}_R) = 1/2 \nonumber \\
P(\text{up}_L | \text{down}_R) = 1 \nonumber \\
P(\text{down}_L | \text{up}_R) = 1 \nonumber
\end{eqnarray}
with $P(\text{up}_L|\text{up}_R) = P(\text{down}_L|\text{down}_R) = 0$.  

On the other hand, the ``linear + cooking'' formulation comes down
instead to the following:
\begin{eqnarray}
P(\text{up}_L, \text{down}_R) = 1/2 \nonumber \\
P(\text{down}_L, \text{up}_R) = 1/2 \nonumber 
\end{eqnarray}
with $P(\text{up}_L, \text{up}_R) = P(\text{down}_L, \text{down}_R) = 0$.  

But these two sets of probability statements are completely
equivalent.  (And the corresponding sets for the case that the
experimenter on the right \emph{doesn't} make a measurement, are even
more trivially equivalent.)  So we are dealing here with two
different-looking recipes which generate the same one stochastic
dynamics for the same set of beables.  There are not two distinct
theories here, only two distinct presentations of the same theory.

It should also be clear that the ``linear + cooking'' theory isn't
really linear at all -- it's just that the non-linearity (which, here,
means dependence of dynamical probabilities for what happens to the
quantum state, on the quantum state itself) is introduced
``at the end''  (in the ``cooking'' process)
as a post-hoc readjustment of the relative likelihoods
for the various things that (otherwise) could have happened.  It may be
that one or the other of the two ways of mathematically formulating
the theory has some advantage in the context of trying to produce
relativistic generalizations from the non-relativistic theory.  For
example, Lorentz covariance may be somewhat more manifest (all other
things being equal) in a ``linear + cooking'' formulation than in a
``non-linear'' formulation.   (See Ref. \cite{bedingham}.)
But this should not be confused with the
two theories being somehow physically different.  Manifest or not, the
two formulations would seem to have to have the same ultimate status
\emph{vis a vis} Lorentz invariance -- and, it would seem, also (in the
non-relativistic context) \emph{vis a vis} nonlocality and the PI/OI
distinction.

\section{GGBF's nonlocality analysis}

The main claim of GGBF, however, is that this is not so:  they allege
that the ``non-linear'' formulation violates PI, while the ``linear 
+ cooking'' formulation instead violates OI.  


We begin with GGBF's straightforwardly correct demonstration that 
the ``linear + cooking'' formulation respects PI.  The proof is just
that the marginal probability, for (say) an ``up'' outcome on the left,
is the same -- namely $1/2$ -- regardless of whether the experimenter on
the right makes a measurement or not.  This is incontestably correct,
and it implies that the ``linear + cooking'' formulation violates 
OI.  (This can be seen directly as well:  for the ``linear + cooking''
theory $P_{|\psi_0\rangle}(\text{up}_L | s_R) = 1/2$ while
$P_{|\psi_0\rangle}(\text{up}_L | s_R, O_R)$ is either
$0$ or $1$ or $1/2$ depending on the realized values of $s_R$ and $O_R$.)

Given all that's been said, then, the puzzle is that GGBF claim also to
prove that the ``non-linear'' formulation violates PI instead.  What
is their argument?

The key to GGBF's argument for this claim is their taking of 
the coin flip on the left as a kind of ``hidden variable''
which, in conjunction with the incoming state vector, determines the
probabilities for the various possible outcomes.  To illustrate, one
may consider for example
the case in which this coin flip on the left happens to give
``heads'' (i.e., the case $H_L$).  
Then the $\lambda$ in Equation (\ref{pi}) subsumes both
$|\psi_0\rangle$ and $H_L$.  
If the experimenter on the
right chooses to make a measurement, we then have that
\begin{equation}
P_{\lambda} (\text{up}_L | \hat{a}) = P_{|\psi_0\rangle} (\text{up}_L | \hat{a},
H_L) = 1/2
\end{equation}
because, while it's given that an experiment happened on the right
so that the state vector which enters on the left will be collapsed,
it is unknown which outcome this collapse produced, so we must average
over the two (equiprobable) possibilities.  

On the other hand, if the experimenter chooses \emph{not} to make a
measurement, we have that
\begin{equation}
P_{\lambda}(\text{up}_L|*) = P_{|\psi_0\rangle} (\text{up}_L | *, H_L) = 1
\end{equation}
since the non-performance of a measurement on the right ensures that
the algorithm generating the outcome on the left will go to the
tie-breaker, in which case the (given) $H_L$ guarantees the outcome
``$\text{up}_L$''.  

Since $1/2 \ne 1$ we have a violation of PI.  That, at least, is the
claim made by GGBF, based on essentially the argument just 
sketched.

But one thing should be immediately clear:  this argument hinges
crucially on taking the outcome of the coin flip on the left as a
hidden variable, i.e., as something that gets brought into the
calculation of probabilities through $\lambda$.  (If this is not done,
the ``non-linear'' formulation, like the ``linear + cooking'' one,
violates OI instead.)    And, arguably, this
is a completely wrong thing to do, because doing it effectively
converts GRW (which is supposed to be an irreducibly stochastic
theory) into a deterministic hidden variable theory.  And we know,
from the beginning, that deterministic theories (which violate Bell's
local causality condition) must necessarily violate PI rather than
OI.

So in that sense the puzzling result is no surprise.  It is just a
consequence of an inappropriate
way of thinking about stochastic theories.  How
should we think about them instead?  We should regard the ``coin
flips'' (more generally, the random numbers that play some role in the
dynamics) not as \emph{beables} (to use Bell's terminology for things that
are physically real), but only as internal aspects of the
algorithm which defines the dynamics \emph{for} the beables.

\section{Discussion}

Let us explore in a bit more detail the consequences of treating the
coin flips as beables in the two formulations of GRW.  To begin with,
it should be clear that if one regards the coin flips the way I think
they should be regarded -- not as beables but as internal aspects of
the algorithm which defines the dynamics \emph{for} the beables --
then both the ``linear + cooking'' and ``non-linear'' formulations of
GRW respect PI and violate OI.  Indeed, understood this way, it is
clear that the two formulations are merely two different mathematical
formulations of the same one physical theory:  they posit the same
beables and mathematically equivalent laws governing the beables.  

As shown just above, however, if one takes the coin flips themselves
as beables for the ``non-linear'' formulation, this becomes in effect
a (non-local) deterministic hidden variable theory.  That it violates
PI rather than OI is then not terribly surprising or interesting.  

One perhaps expects the same thing to occur with the ``linear +
cooking'' formulation, but here there is something that is surprising
and interesting.  Recall that, in the ``linear + cooking'' formulation, the
coin flip outcomes simply determine the corresponding measurement
outcomes:  $H_L \rightarrow \text{up}_L$, and so on.  So if we
consider the outcomes of the coin flips as given hidden variables, we
have for example that, for $\lambda = ( |\psi_0\rangle, H_L)$, 
\begin{equation}
P_{\lambda} (\text{up}_L | s_R ) = 1
\end{equation}
and that
\begin{equation}
P_{\lambda} (\text{up}_L | s_R, O_R) =1
\end{equation}
so that OI is respected.  But 
\begin{equation}
P_{\lambda} (\text{up}_L | \hat{a} ) = P_{\lambda} (\text{up}_L | * )
= 1 
\end{equation}
so PI is respected as well!  What is going on here?  

The point is that, in the ``linear + cooking'' formulation, the
non-locality is in the part of the dynamics that brings about some
particular set of coin flip outcomes.  (Recall that the coin flip
outcomes are, after cooking, \emph{correlated} despite their 
association with spacelike
separated spacetime regions.)   So if one regards these coin
flip outcomes as simply given, the remaining dynamics (which in effect
just has the experimental outcomes being read off from the coin flips)
is completely local.  It makes the theory into a kind of conspiracy
theory, in that the otherwise-puzzling correlations between
measurement outcomes are written in advance
into some given (and so unexplained) correlations in the hidden
variables.  But still, the theory is local.\footnote{And note that, in
  this way of thinking about the theory, the conspiracy goes very
  deep.  The probabilities for various coin flip outcomes depend
  (through the cooking) on the way the wave function evolves.  And
  that in turn depends on the realized values of external fields,
  e.g., whether $\hat{a}$ or $*$ obtains.  Thus, the notion of
  given, previously-assigned values for all the coin flip outcomes
  precludes the possibility of free-will choices or, equivalently
  here, free variables (such as ``external fields'').  This is not,
  therefore, a way of thinking about GRW that I think one should take
  seriously.  It is in the category of theories Bell described as
  ``superdeterministic''.}

To summarize the possibilities:  by fiddling around with (a) 
which elements of a theory are granted ``beable status'' \cite[p. 53]{bell}
and (b) whether certain beables are regarded as simply given, or instead as
unfolding in some dynamical process, GRW can be variously 
understood as a non-local stochastic theory (which violates OI but 
respects PI), a non-local deterministic theory (which respects OI and
violates PI), or a local conspiracy theory (which respects both OI and
PI).  


From the point of view of Bell's careful formulation of local
causality for candidate theories, none of this is terribly
surprising.  Indeed, Bell's locality criterion is ``in terms of local
beables'' \cite[p. 53]{bell} so it stands to reason that theories with
different beables might have different standings with respect to local
causality.\footnote{I made a point earlier of insisting that the
  ``non-linear'' and ``linear + cooking'' formulations were not
  different theories, on the grounds that they posited the same
  beables (namely just the wave function) and also mathematically
  equivalent laws for the dynamical evolution of those beables.  The
  point here is that, by granting ``beable status'' to some additional
  elements in the theory (such as our coin flips), one \emph{does}
  produce a genuinely new and distinct theory, and the question of
  \emph{its} status vis a vis locality, OI, PI, etc., must be
  addressed anew.  As a parallel example, consider orthodox
  one-particle quantum mechanics.  This is a non-local theory, as
  shown most easily by the ``Einstein's Boxes'' argument. \cite{boxes}
  But by adding an additional beable -- the actual position of a
  particle within the wave -- one has a new theory which, at least in
  the Einstein's Boxes scenario, is perfectly local.  All of this
  illustrates the importance of recognizing that Bell's local
  causality condition applies to \emph{candidate theories} and is ``in
  terms of local beables''.}

Note also the connection between the mistake in GGBF's argument,
analyzed above, and the ongoing controversy about the 
validity and implications of Conway and Kochen's so-called Free Will
Theorem.  
Conway and Kochen argue,
wrongly, that a stochastic theory will necessarily have the same
status (\emph{vis a vis} local causality and/or relativity) as the
deterministic theory it would be converted into by taking the
background random numbers as beables.  As they put this claim, we can 
\begin{quote}
``let the stochastic element ... be a sequence of random numbers (not
all of which need be used by both particles).  Although these might
only be generated as needed, it will plainly make no difference to let
them be given in advance.'' \cite{ck}
\end{quote}
That in fact it \emph{does} make a difference has been pointed out
lucidly in Ref. \cite{fwtc}.  (The authors show that, in their
terminology, Tumulka's rGRWf theory is ``effectively causal'' -- whereas the
deterministic theory it would be converted into by taking all the
random numbers as ``given in advance'' \emph{isn't}.)

It is hoped that the analysis of the (simpler) example of this same
error -- discussed in the current paper --
will help clarify and underscore this important lesson.

\end{document}